\documentclass[cernpreprint, atlasdraft=false, texlive=2025, UKenglish, coverpage=false, orcidlogo=true, texmf, orcidlogo]{atlasdoc}
\usepackage{atlaspackage}
\usepackage[backref=false]{atlasbiblatex}

\usepackage[figurename=Figure]{caption}

\usepackage{atlascontribute}

\DeclareOldFontCommand{\bf}{\normalfont\bfseries}{\mathbf}
\usepackage[unit=true, hepparticle=true, hepprocess=true, jetetmiss=true, misc=true]{atlasphysics}
\graphicspath{{logos/}{figures/}}

\usepackage{ANA-TOPQ-2022-22-PAPER-defs}

\PreprintIdNumber{CERN-EP-2023-183}
\AtlasJournalRef{Phys. Lett. B 854 (2024) 138726}
\AtlasDOI{10.1016/j.physletb.2024.138726}

\AtlasTitle{Measurement of $t$-channel single-top-quark production in $pp$ collisions at $\sqrt{s} = 5.02$ TeV with the ATLAS detector}
\AtlasJournal{\PLB}

\AtlasAbstract{%
The observation of the electroweak production of single-top-quarks is made using 255 $\text{pb}^{-1}$ of proton-proton collision data recorded at $\sqrt{s} = 5.02$ TeV with the ATLAS detector at the Large Hadron Collider.
An event selection is used to identify single-top-quark candidates arising from $t$-channel production with the top quark decaying semi-leptonically. Events passing the selection are then used to measure the inclusive cross-section for the combined production of  single-top-quarks
and antiquarks, $\sigma({tq + \bar{t}q})$, and the ratio $R_t$ between these two.
They are measured to be $\sigma({tq + \bar{t}q})  = 27.1^{+4.4}_{-4.1} (\text{stat.}) ^{+4.4}_{-3.7} (\text{syst.})$ pb and $R_t = 2.73^{+1.43}_{-0.82} (\text{stat.}) ^{+1.01}_{-0.29} (\text{syst.})$. The individual single-top-quark ($t q$) and single-top-antiquark ($\bar{t} q$) production cross-sections are measured to be $\sigma(t q) = 19.8^{+3.9}_{-3.1} (\text{stat.}) ^{+2.9}_{-2.2} (\text{syst.})$ pb and $\sigma(\bar{t}q) = 7.3^{+3.2}_{-2.1} (\text{stat.}) ^{+2.8}_{-1.5} (\text{syst.})$ pb.
All measurements are in good agreement with the Standard Model predictions.
}

\author{The ATLAS Collaboration}

\hypersetup{pdftitle={ATLAS document},pdfauthor={The ATLAS Collaboration}}

\usepackage{stackengine}

\begin{document}

\maketitle

%
%

%
%
%
%
%
\section{Introduction}
Top quarks can be produced in pairs via the strong interaction or singly via the
electroweak interaction at hadron colliders~\cite{Willenbrock:1986cr}.
Single-top-quark production was first observed in anti-proton--proton collisions at the
Tevatron~\cite{CDF, DO, CDF:2015gsg} and since then it has been extensively studied as a window into the
properties of the top quark itself~\cite{Schwienhorst}.
These include studies of the unitarity of the
Cabibbo-Kobayashi-Maskawa matrix ~\cite{PhysRevLett.10.531, PTP.49.652, ATLAS:2019hhu, CDF:2015gsg},
tests of higher-order corrections from quantum chromodynamics (QCD)~\cite{PhysRevLett.102.182003}, and
constraints on the parton distribution functions (PDF) of the proton~\cite{PhysRevD.104.094042}.

\begin{figure}[h]
\centering
\subfloat[]{
\includegraphics[width=0.5\linewidth]{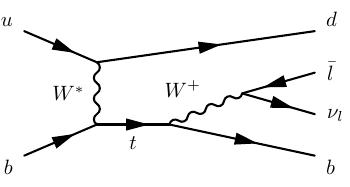}
}
\subfloat[]{
\includegraphics[width=0.5\linewidth]{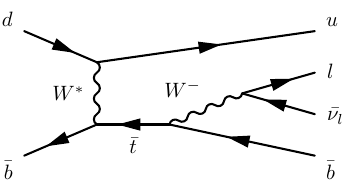}
}

\caption{Feynman diagrams at leading-order in QCD for (a) single-top-quark production and (b) single-top-antiquark production via the $t$-channel exchange of a virtual $W$ boson ($W^*$). The diagrams include the leptonic decay of the top quark and top antiquark.}
\label{fig:diagrams}
\end{figure}

Electroweak theory predicts three primary mechanisms for single-top-quark production in proton--proton (\pp) collisions at the Large Hadron Collider (LHC): $t$-channel, $s$-channel, and $tW$ (or $W$-associated) production.
The $t$-channel mechanism for single-top-quark production, shown in
Figure~\ref{fig:diagrams}, has the largest cross-section of the three mechanisms and also the
final-state topology with the highest signal-to-background ratio, due to the presence of a light-quark jet recoiling against the top quark that assists in identifying this topology.
The $t$-channel production cross-section, \xtchan, was measured by
both the ATLAS and CMS collaborations at centre-of-mass energies of $\rts = 7$ \TeV~\cite{TOPQ-2011-14,TOPQ-2012-21,CMS-TOP-10-008,CMS-TOP-11-021}, $\rts = 8$ \TeV~\cite{TOPQ-2015-05, CMS-TOP-12-038} and $\rts = 13$ \TeV~\cite{TOPQ-2015-15, CMS-TOP-16-003, CMS-TOP-17-011}.

The observation of $t$-channel single-top-quark production and a measurement of its cross-section at \rtsf\ is reported using \SI{255}{$\text{pb}^{-1}$}\ of \pp\ collision data collected with the ATLAS detector.
The analysis includes selection criteria to isolate the $t$-channel topology from the
Standard Model (SM) backgrounds in the leptonic-decay channels of the \Wboson\ boson
($t \rightarrow e \nu b$, $t \rightarrow \mu \nu b$, and $t \rightarrow \tau \nu b$ with
leptonic $\tau$-lepton decays).
Separate measurements of the single-top-quark final states with a top quark and a top-antiquark
are used to measure the CKM matrix element $V_{tb}$.
This measurement at a lower centre-of-mass energy than other proton-proton results provides an
independent test of the SM, with different levels of backgrounds and instrumental uncertainties.

The backgrounds arise from the production of $W$ bosons in association with jets (\Wjets), top-quark pair (\ttbar) production, and either misidentified or non-prompt leptons, wherein a jet is mis-reconstructed as an electron, or a heavy-flavour quark that decays into a muon which satisfies the selection criteria.
The two subleading single-top-quark production mechanisms, the production of $Z$ bosons in association with jets (\Zjets), and diboson ($WW$, $WZ$, and $ZZ$) processes are additional minor backgrounds.
Some of these processes were measured by the ATLAS and CMS collaborations~\cite{ATLAS:2022jbj, CMS-TOP-20-004, CMS-SMP-20-012} at $\rts = 5.02$ \TeV.

\section{The ATLAS detector}
The ATLAS detector~\cite{PERF-2007-01} at the LHC is a multipurpose particle-physics detector with a cylindrical geometry.\footnote{ATLAS uses a right-handed coordinate system with its
origin at the nominal interaction point (IP) in the centre of the detector and the \(z\)-axis along the beam pipe.
The \(x\)-axis points from the IP to the centre of the LHC ring, and the \(y\)-axis points upwards.
Cylindrical coordinates \((r,\phi)\) are used in the transverse plane, \(\phi\) being the azimuthal angle around the \(z\)-axis.
The pseudorapidity is defined in terms of the polar angle \(\theta\) as \(\eta = -\ln \tan(\theta/2)\). Momentum in the transverse plane is denoted by \pT.}
It consists of an inner tracking detector (ID) surrounded by a 2~T superconducting solenoid, sampling electromagnetic (EM) and hadronic calorimeters, and
a muon spectrometer (MS) incorporating three large superconducting toroid magnets. A two-level trigger system is used to select events for storage.

An extensive software suite~\cite{ATL-SOFT-PUB-2021-001} is used in the reconstruction and analysis of real and simulated data, in detector operations, and in the trigger and data acquisition systems of the experiment.
Events used in this measurement were collected using single-electron or single-muon triggers~\cite{TRIG-2016-01}.

\section{Event selection}
The measurement is performed on \SI{255}{$\text{pb}^{-1}$}\ of \pp\ collision data collected at \sxut\ with the ATLAS detector, after applying data quality requirements~\cite{DAPR-2018-01}. Most triggered events
also included signals from additional inelastic $pp$ collisions in the same bunch crossing, referred to as
pile-up. This special data sample was taken in November 2017 under low-pile-up conditions where the mean number of inelastic \pp\ collisions per bunch crossing was ${\approx}2$~\cite{ATLAS-CONF-2020-023}.
This analysis follows a measurement of the \ttbar\ cross-section at the
same centre-of-mass energy~\cite{ATLAS:2022jbj}\ and employs the same algorithms to calibrate the data.

The selected \pp\ interaction vertex is the one with the highest $p_{T}^{2}$ sum of matched tracks with at least two matched tracks required to have a transverse momentum $\pt > 0.5$~\GeV.
Electron candidates are reconstructed from localised clusters of energy deposits in the EM calorimeter that are matched with tracks found in the ID\@. Muon candidates are reconstructed by combining and matching tracks reconstructed in the ID with tracks or track segments found in the MS\@.
Electrons (muons) must have $ \pt > 18 $\,\GeV\ and be reconstructed in a
pseudorapidity range of $|\eta| < 2.47$ ($|\eta| < 2.5$);
electrons in the range of $1.37 < |\eta| < 1.52$\ are excluded.
To ensure that selected leptons originate from the primary vertex, their tracks are required to have $|d_0/\sigma_{d_0}| < 5~(3)$ for electrons (muons) and $|z_0 \sin \theta|< 0.5$~mm for both the lepton flavours.
Here $d_0$ and $\sigma_{d_0}$ are the transverse impact parameter and its uncertainty, while $z_0$ is the longitudinal impact parameter.
Electrons are required to satisfy the `medium' likelihood-based identification criterion defined in Ref.~\cite{EGAM-2018-01} while muons must satisfy the `medium' cut-based identification criterion defined in Ref.~\cite{MUON-2018-03}.
Leptons likely to originate from light-hadron decays or heavy-flavour decays are rejected by applying a `tight' isolation requirement as defined in Ref.~\cite{TOPQ-2015-09}.
Particle-flow jets are reconstructed from tracks in the ID and topological clusters of
calorimeter energy deposits~\cite{PERF-2015-09} using the
\antikt\ algorithm~\cite{Cacciari:2008gp,Fastjet}\ with a radius parameter $R=0.4$.
These jets are calibrated according to the standard calibration used for \sxyt\ high-pile-up data~\cite{JETM-2018-05}.
An additional correction to the jet-energy scale, of 2\%-12\%, is derived using the technique of balancing the \pt of $Z$\,+\,jet events and applied to data. This additional correction is used
to account for the modified calorimeter response in the low-pile-up data sample~\cite{ATLAS:2022jbj}.
Jet candidates are required to have $\pt > 23$\,\GeV\ and $|\eta| < 4.0$. Jets with $\pt < 60$\,\GeV\ and $|\eta|<2.4$ are subject to additional pile-up rejection criteria using a multivariate jet-vertex tagger~\cite{PERF-2014-03}.

Jets originating from long-lived $b$-hadrons are identified using the DL1r algorithm~\cite{ATLAS:2022qxm}, a multivariate discriminant based on deep-learning techniques using information from track impact parameters and reconstructed secondary vertices. A working point with 60\% efficiency for tagging $b$-quark jets from top-quark decays in simulated \ttbar\ events is used. At this working point, the tagger has rejection factors of 30 against charm jets and 1200 against light-quark jets. Jets passing this requirement are denoted as $b$-tagged jets.

The missing transverse momentum, whose magnitude is denoted by \etmiss, is reconstructed as the negative vector sum of the transverse momenta of all identified physics objects (electrons, muons, and jets), together with a `soft term' built from all tracks matched with the reconstructed primary vertex but not with any of the identified physics objects~\cite{PERF-2016-07}.

Selected events are required to have exactly one electron or muon candidate, and exactly two jets.
Exactly one of the jets must be $b$-tagged and be in the central region $|\eta|<2.5$.
Since the spectator-quark jet tends to be produced in the forward direction in the $t$-channel process, the pseudorapidity of the untagged jet must satisfy $1.5< |\eta| <4.0$.
To reduce contributions from the \ttbar process, the pseudorapidity separation between the untagged jet and the $b$-tagged jet is required to be ${>}1.5$.
A \Wboson\ boson candidate is identified by the electron or muon candidate and the
missing transverse momentum.
To suppress contributions from misidentified leptons, the following cuts on $\met$ and the transverse mass of the \Wboson boson\footnote{$\mtw=\sqrt{2p^{\ell}_{\mathrm T}\met\,(1-\cos\phi)}$, where $p^{\ell}_\mathrm{T}$ is the transverse momentum of the charged lepton and $\phi$ is the opening azimuthal angle between the charged lepton and the missing transverse momentum.} (\mtw) are applied: $\mtw > 35$\,\GeV, $\met>15$\,\GeV, and $\met + \mtw > 70$\,\GeV.
To increase the purity of the signal events, the \HT\ is required to be greater than 185 GeV, where \HT\ is defined as the scalar sum of the \pT\ of the jets, the \pT\ of the lepton, and \etmiss\,.
A top-quark candidate is reconstructed in each event by combining the $b$-tagged jet with a $W$ boson candidate. The latter is kinematically reconstructed by imposing the $W$ boson mass as a kinematic constraint on the sum of the electron or muon candidate
and the missing transverse momentum, leading to a quadratic equation in the longitudinal neutrino momentum, $p_{\nu,z}$. When two solutions are found, the one that gives a reconstructed top-quark mass closest to the on-shell mass (172.5 \GeV) is chosen. For complex solutions only the real component is considered, which results in a reconstructed mass greater than the $W$\ boson mass. This phenomena is observed in approximately 30\% of the events.
To further suppress contributions from processes not involving top quarks, the invariant mass of the lepton and the $b$-tagged jet must be ${<}165$~\GeV, the invariant mass of the reconstructed $W$ boson must be ${<}102$~\GeV, and the invariant mass of the reconstructed top quark must be between $140$~\GeV\ and
$225$~\GeV.

\section{Modelling and theoretical predictions}
The \MCFM\ program~\cite{Campbell_2010, campbell2015multithreaded, Campbell:2019}\ employing
the \NNPDF[3.0nlo] PDFs~\cite{Ball:2012cx} was used to calculate predictions for the single-top-quark ($tq$) and single-top-antiquark ($\bar{t}q$) production
cross-sections in the $t$-channel at next-to-next-to-leading-order (NNLO) in QCD for \pp\ collisions at $\sqrt{s} = 5.02$~\TeV.
The predicted values
are $\sigma(tq)=20.3^{+0.5}_{-0.4}$\,pb,
$\sigma(\bar{t}q)=10.0^{+0.2}_{-0.3}$\,pb,
$\xtchan = 30.3^{+0.7}_{-0.5}$\,pb. The ratio of $\sigma(t q)$ to $\sigma(\bar{t}q)$, $R_t$ is predicted to be
$R_t = 2.03^{+0.06}_{-0.07}$. All predictions are
assuming a top-quark mass of $172.5$\,\GeV.
The quoted uncertainties include contributions from the choice of renormalisation
scale \muR\ and the factorisation scale \muF, the uncertainty in the PDFs, and uncertainty in the value of \alphas.

Monte Carlo (MC) simulated samples are used to model the single-top-quark $t$-channel signal process and contributions from other physics processes with prompt leptons.
The signal was simulated using the next-to-leading-order (NLO) in QCD MC event generator \POWHEGBOX[v2]~\cite{Frederix:2012dh} with the four-flavour scheme. The
parton showering, hadronisation, and the underlying event was
modelled using the
\PYTHIA[8.2]~\cite{pythia8} program with
the A14 tune~\cite{ATL-PHYS-PUB-2014-021}.
The \POWHEGBOX[v2] program also was used to simulate \ttbar, single-top-quark $s$- and $tW$-channel backgrounds. All these samples use \PYTHIA[8.2]\ with
the \NNPDF[3.0nlo] PDFs and
the A14 tune as the parton-shower and hadronisation models.
The \ttbar\ background is normalised using an NNLO cross-section, while the $tW$ background is
normalised using an approximate NNLO cross-section including the resummation of next-to-next-to-leading logarithmic soft-gluon terms~\cite{toppp, Kidonakis:2021vob}.
The $s$-channel sample is normalised to the generator-level NLO cross-section prediction.

The \Zjets\ and \Wjets\ events were simulated with the \SHERPA[2.2] generator~\cite{sherpa22} using NLO matrix elements with up to two partons, and leading-order (LO) matrix elements for up to four partons, and normalised using an NNLO cross-section prediction~\cite{zxsecnnlo}.
The \NNPDF[3.0nnlo]\ PDF sets~\cite{Ball:2014uwa} were used for all \Zjets\ and \Wjets\ simulated samples.
The smaller backgrounds from diboson production with additional jets were simulated using the \SHERPA[2.1]\ generator with the \CT[10] PDF set~\cite{cttenpdf}.

All generated events underwent a full simulation of the ATLAS detector response based on the $\GEANT$~\cite{Agostinelli:2002hh} framework.
The effects of pile-up are included in the simulation. To  improve the
agreement with the response observed in data, small corrections derived from comparisons of data and simulation at both \sxut\ and \sxyt\ were applied as scale factors to the simulated lepton-trigger and
lepton-reconstruction efficiencies.

The backgrounds arising from the non-prompt and misidentified leptons were determined using the `matrix method' technique~\cite{ATLAS:2022swp}. Events were selected using looser isolation or identification requirements for the lepton and were then weighted according to the efficiencies for both the prompt and background (misidentified and non-prompt) leptons to pass the tighter baseline selection. The method was validated by comparing predictions with data in dedicated validation regions with a larger fraction of misidentified-lepton candidates than expected in the analysis regions. Good agreement between data and the prediction in these validation regions was found.

\section{Signal and cross-section extraction}
Boosted decision trees (BDT) are used to enhance the separation between signal and background.
A single BDT is trained using nine input variables that include information about object kinematics, variables based on combinations of four-vectors, and the global event topology.
The BDT is trained with the XGBoost~\cite{XGBoost} package using MC signal, MC background, and the data-driven misidentified-lepton background events.
The variables with the highest discriminating power between signal and background are the total scalar sum of the transverse momentum from all objects in an event (\HT), and the magnitude of the  \pt\ difference  between the reconstructed \Wboson\ boson and the four-vector sum of the untagged and $b$-tagged jet (\AbsDPtWub).

A three-fold cross-validation procedure is used to produce the final discriminant that is calculated
for the observed data and the predictions.
The sample is divided into three subsamples while
the training is performed on a pair of the subsamples and tested against the third.
The procedure is iterated with three different pairings to produce a combined BDT discriminant,
which is applied to the data events.
A binned profile-likelihood fit of the sum of the BDT response distribution for signal and background MC samples to
the observed BDT response distribution is performed.

\begin{figure}[t]
\begin{center}
\subfloat[ ]{
\includegraphics[width=0.45\linewidth]{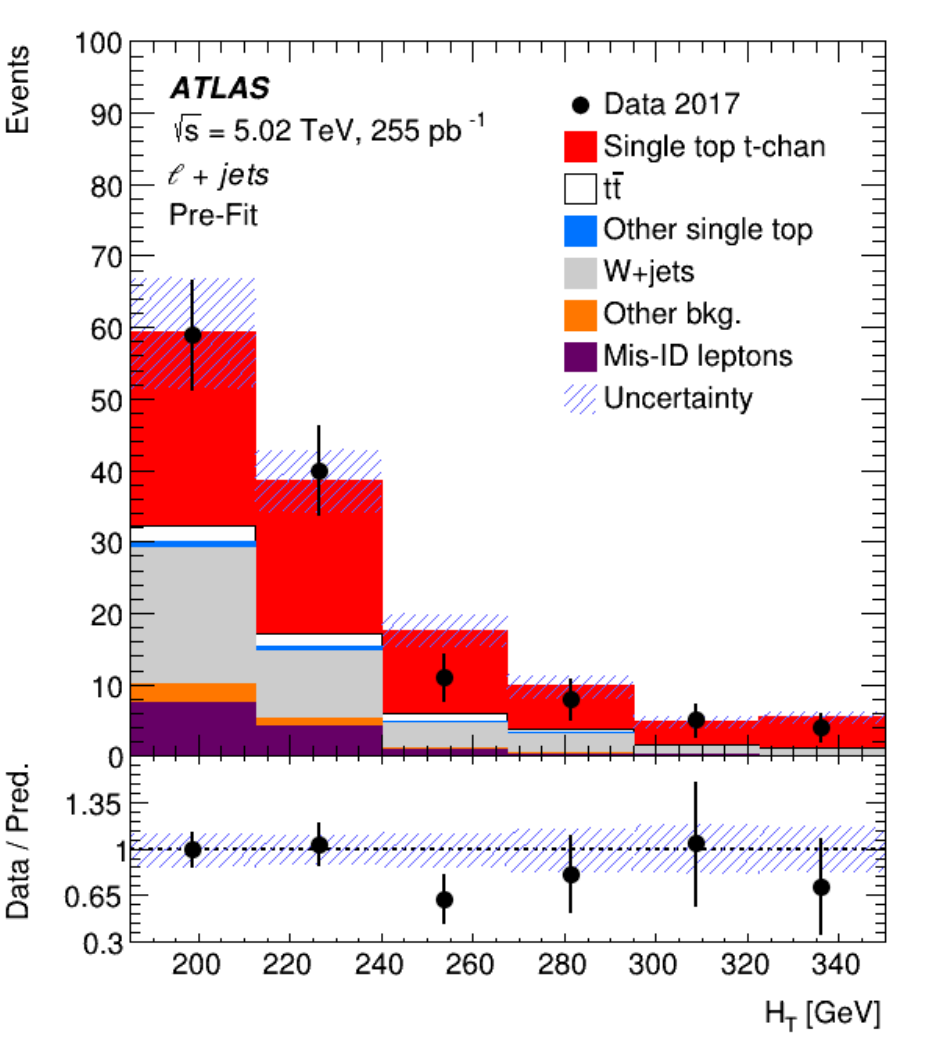}
}
\quad
\subfloat[ ]{
\includegraphics[width=0.45\linewidth]{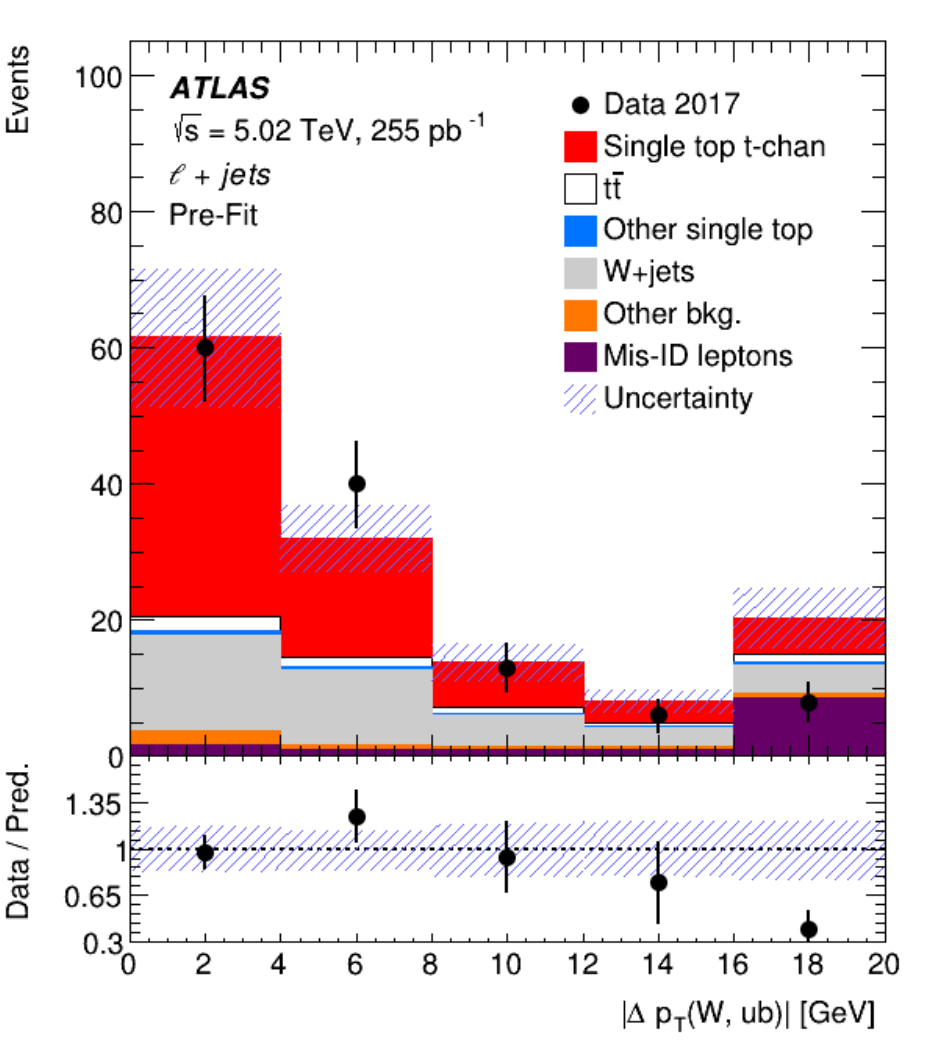}
}
\caption{The pre-fit (a) \HT\ and (b) \AbsDPtWub\ distributions for data (dots) in the inclusive $\ell$+jets channel. The MC simulation of the signal (red histograms) and various backgrounds (represented by histograms of different colours) are also included. The error bars on the dots represent the statistical uncertainty on the data while the blue cross-hatched lines correspond to the total uncertainties on the prediction.
The lower panels show the ratio of the data and the prediction, along with the uncertainty in the ratio. The last bin includes any event overflows. The $\chi^2$/degrees of freedom are evaluated to be 2.5/6 for (a) and 5.9/5 for (b).
}
\label{fig:PreFit}
\end{center}
\end{figure}

The  $\sigma(t q)$ and $\sigma(\bar{t}q)$ cross-sections are determined by dividing the
event sample into a top-quark and top-antiquark subsample defined by the charge of the
reconstructed lepton,
$\ell^{+}$+jets and $\ell^{-}$+jets, respectively.
These cross-sections are parameterised as
functions of \xtchan\ and $R_t$, and
are determined by the fit to the observed BDT distributions in these two subsamples.
The \HT\ and \AbsDPtWub\ distributions for the inclusive $\ell$+jets sample (the sum of the $\ell^{+}$+jets and $\ell^{-}$+jets regions) are shown in Figure~\ref{fig:PreFit}, along
with the predictions from the signal and background model. The predicted distributions show the signal assuming the SM single-top-quark $t$-channel
production cross-section and the estimated backgrounds for the two subsamples
before the fit is performed (pre-fit).
The predicted shapes of the distributions for this variable and the others used in the analysis are
found to be in good agreement with the observed distributions.

\section{Systematic uncertainties}
Systematic uncertainties are included in the likelihood fit as nuisance parameters
constrained by Gaussian probability density functions. Correlations between systematic uncertainties arising from common sources are maintained across processes and bins in the two regions.

Uncertainties arising from the modelling of signal- and background-related processes are evaluated using alternative MC samples.
The \singletop\ signal has uncertainties arising from the matrix-element matching, the
parton-shower and hadronisation model, initial and final-state radiation (ISR and FSR),
the \muR\ and \muF\ scales in the matrix element, and the
proton PDFs~\cite{ATL-PHYS-PUB-2014-021, Butterworth}.
Two alternative $t$-channel samples were used to evaluate the matching uncertainty and
uncertainties in the parton-shower model: one was generated
using the nominal matrix-element model and
the \texttt{pTHard} parameter in \PYTHIA[8]\ changed from zero to one, and the second generated
using the \POWHER[7.1.6] generator.

The \Wjets\ modelling uncertainties are evaluated by first splitting the MC generated \Wjets\ background into three categories $W+{\ge}1b$, $W+{\ge}1c$,
and $W+{\ge}1l$ (light quarks or gluons) based on the flavour of additional jets in the event, with fractions in each category of 57\%, 39\%, and 4\%, respectively.
The uncertainty in each fraction is calculated by adding in quadrature a 24\%\ uncertainty for each successive jet in the event using  Berends scaling~\cite{Berends:1990ax}.
Thus, a conservative \Wjets\ normalisation uncertainty of 34\% is used for all categories of \Wjets\ background in lieu of generator comparisons since alternate MC event samples were not available at $\rts = 5.02$ TeV. The \muR\ and \muF\ scales are varied by factors of $1/2$\ and 2 and an envelope built from six possible variations is used to estimate additional shape uncertainties for the three categories of \Wjets\ background.

The uncertainty in the cross-section of the \ttbar\ background is taken to be $^{+7.5}_{-7.7}$\%~\cite{toppp, Kidonakis:2021vob}, while modelling uncertainties are estimated by using the techniques in Ref.~\cite{ATLAS:2022jbj}.
Uncertainties in the cross-sections for the $tW$ and $s$-channel single-top-quark background processes are taken to be $^{+5.6}_{-5.9}\%$ and 9.5\%, respectively~\cite{hathor, hathor15}. Uncertainties related to parton-shower, modelling of ISR and FSR, and the \muR\ and \muF\ scales for these three background processes are included in the fit. The PDF uncertainties are also included for the \ttbar\ background.
A conservative 50\% uncertainty in the cross-section, acceptance, and modelling of the \Zjets\ and diboson backgrounds is applied~\cite{Campbell_2010}. The fit also includes a 50\% uncertainty in the normalisation of the misidentified-lepton background estimate that is determined by comparing different parameterisations
and selections for extracting the lepton efficiencies used in the matrix method. The uncertainty is separated according to the flavour of the misidentified lepton.

Instrumental systematic uncertainties related to the lepton trigger efficiency~\cite{TRIG-2018-01, TRIG-2018-05}, reconstruction, isolation and identification~\cite{PERF-2017-01, EGAM-2018-01, MUON-2018-03}, lepton energy scale and resolution~\cite{PERF-2017-03}, and jet-energy scale and resolution~\cite{JETM-2018-05} are measured in the \sxut\ data sample or taken from high-pile-up \sxyt\ data with additional uncertainties to account for the extrapolation to the low-pile-up\ \sxut\ data.
Due to the additional correction to the jet-energy scale specific to this low-pile-up data sample,
statistical and modelling uncertainties of 1\%--2\%\ on the jet-energy scale arising
from the correction are incorporated into the systematic uncertainties in the fit. The uncertainty on the $b$-tagging efficiency is measured to be $\sim 1\%$~\cite{ATLAS:2022qxm}.
Additional uncertainties arise from jet-vertex tagging~\cite{PERF-2014-03} and modelling of \met~\cite{PERF-2016-07}. The uncertainty in the integrated luminosity is 1.0\% and the uncertainty in the beam energy is negligible.

\section{Results}
\begin{figure}[t]
\begin{center}
\subfloat[ ]{
\includegraphics[width=0.45\linewidth]{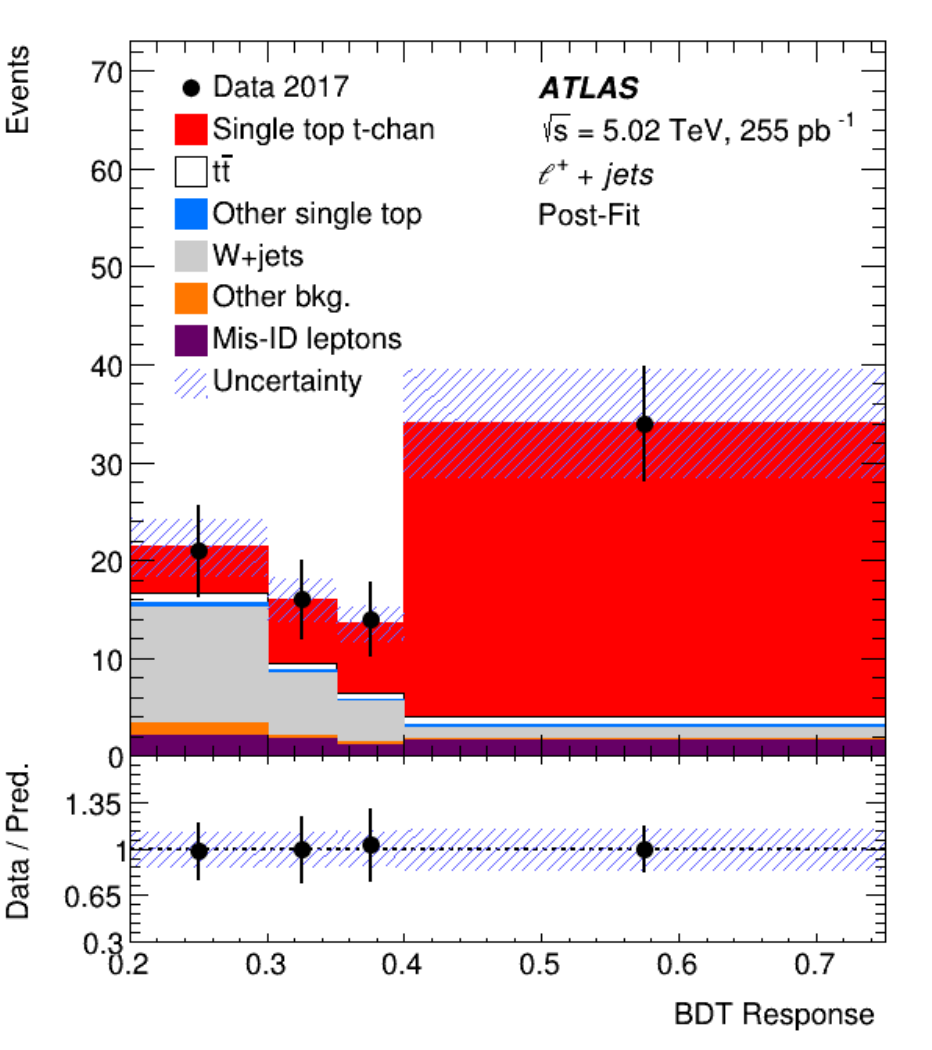}
}
\quad
\subfloat[ ]{
\includegraphics[width=0.45\linewidth]{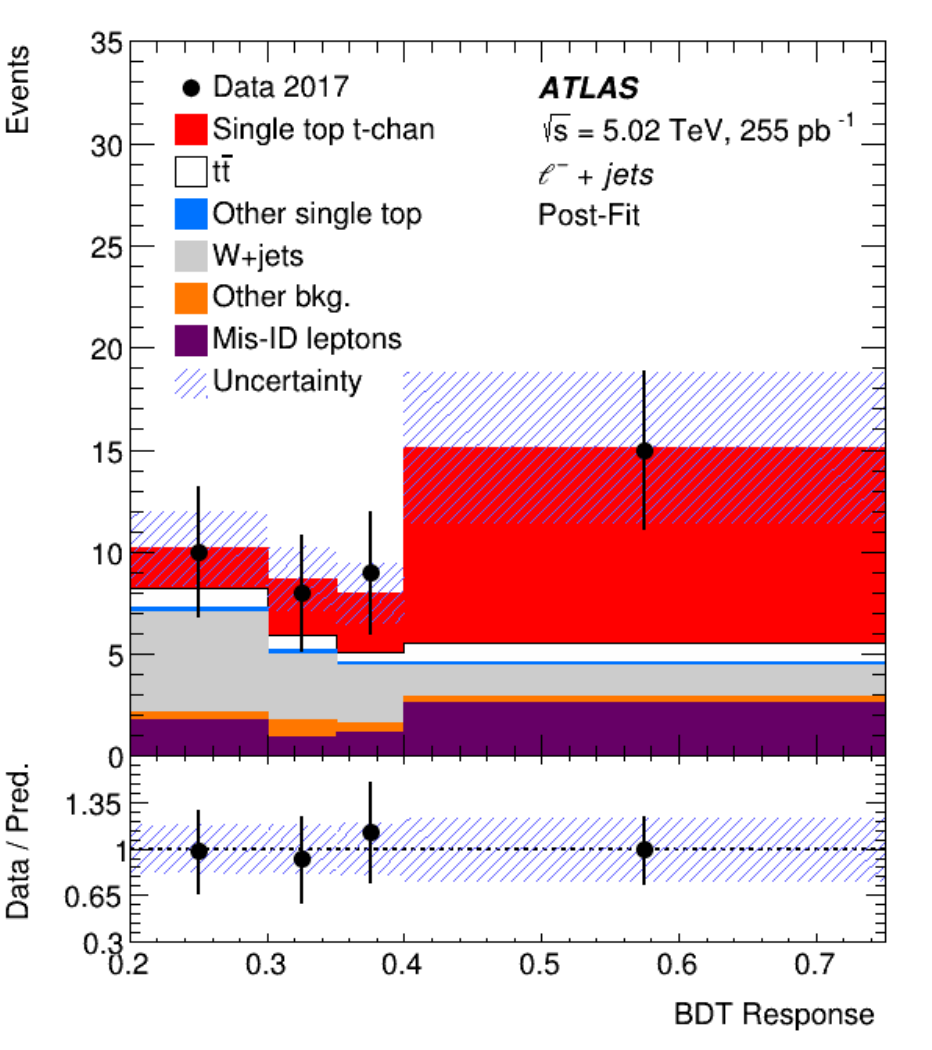}
}

\caption{The post-fit BDT response distribution for data (dots) in the (a) $\ell^{+}$+jets and (b) $\ell^{-}$+jets channels. The MC simulation of the signal (red histograms) and various backgrounds (represented by histograms of different colours) are also included. The error bars on the dots represent the statistical uncertainty on the data while the blue cross-hatched lines correspond to the total uncertainties on the prediction.
The lower panels show the ratio of the data and the prediction, along with the uncertainty in the ratio. Bins with BDT response values outside of the bins shown are empty.
}
\label{fig:BDT}
\end{center}
\end{figure}

The results of the profile likelihood fit are $\xtchan = \Writetcx$~pb and $R_t = \WriteRt$. The Pearson correlation coefficient between $\xtchan$ and $R_t$ is measured to be $\FullSystCorrBetweenTchanAndRT$ and all other correlations are found to be small.
Table~\ref{tab:yields} shows the fitted signal- and background-event yields (post-fit) and the observed yield in the $\ell^{+}$+jets and $\ell^{-}$+jets regions.
The fit to the BDT response distributions is shown in Figure~\ref{fig:BDT}.
The pulls of all nuisance parameters in the fit are found to be within 0.2 standard deviations
of their input values.
\begin{table}[t]
\caption{Number of post-fit signal, background and observed data events in the $\ell^{+}$+jets and $\ell^{-}$+jets regions. The `Other single top' category contains the $tW$ associated production and $s$-channel contributions. The uncertainties in the signal and background yields include the statistical uncertainties, all systematic uncertainties, and the correlations between them.
The uncertainty in the total prediction includes correlations between all systematic uncertainties and thus does not equal the sum in quadrature from the individual components.
In this table, the uncertainties have been symmetrised, but full asymmetric uncertainties are used to obtain the final results.
}
\begin{center}
\begin{tabular}{ l  r@{$\,\pm\,$}l r@{$\,\pm\,$}l }
\toprule
Source               & \multicolumn{4}{c}{Number of events}   \\
& \multicolumn{2}{c}{$\ell^{+}$+jets} & \multicolumn{2}{c}{$\ell^{-}$+jets}  \\

\midrule
\singletop           & 49 & 9 & 17 & 8  \\
$W$ + jets             & 23 & 5 & 12 & 3 \\
Misidentified leptons & 7 & 3 & 7 & 3 \\
$\ttbar$             & 3  & 0.5 & 3 & 0.5   \\
\Zjets\ and diboson  &  2 & 1 & 2 & 1  \\
Other single-top-quark production     & 1  & 0.2 & 1 & 0.5 \\
\midrule
Total predicted     & 85 &  9 & 42 & 7   \\
\midrule
Data                 & \multicolumn{1}{l}{85} & & \multicolumn{1}{l}{42}   \\
\bottomrule
\end{tabular}
\label{tab:yields}
\end{center}
\end{table}

Using the asymptotic approximation~\cite{Cowan:2010js}, the background-only
hypothesis is rejected with an observed (expected) significance of \SignificanceObs\ (\SignificanceExp) standard deviations.

The statistical uncertainty from the sample size is the largest contribution to the
total uncertainty in the single-top-quark $t$-channel production cross-section,
followed by a $8.5$\% uncertainty contribution from signal-modelling
and a $6.3$\% contribution from modelling uncertainties in the misidentified-lepton background.
The signal-modelling uncertainty is dominated by the uncertainty in the choice of parton-shower
and hadronisation models.
Table~\ref{tab:SysImpactData} shows the breakdown of the sources of
uncertainty in the cross-section and cross-section ratio measurements.
The squared uncertainties are calculated by fixing the set of nuisance parameters corresponding
to a category, repeating the fit, and taking a difference between the squares of the resulting
uncertainty and the total uncertainty of the nominal fit.
The total uncertainty is the sum in quadrature of the total systematic uncertainty and the data's statistical uncertainty.

\begin{table}[tp]
\caption{Sources of uncertainty for measurements of \xtchan\ and $R_t$ at \sxut.
The systematic uncertainties for both the values do not add up in quadrature to the total
systematic uncertainty because of correlations in the fit parameters.
In this table, the uncertainties have been symmetrised, but full asymmetric uncertainties are used to obtain the final results.
}
\centering
\begin{tabular}{ l  c  c  }
\toprule
Category & $\delta \xtchan / \xtchan[\%]$ & $\delta R_t / R_t [\%]$ \\
\toprule
Single-top quark signal modelling            &  8.6  &  4.1 \\
Parton distribution functions & 0.5   &  0.8 \\
\midrule
Misidentified leptons background  & 6.3 &  11.1      \\
$W + \geq1b$ jets modelling     & 3.9 &   4.4    \\
$W + \geq1c$ jets modelling     & 2.7 &    3.4  \\
$Z$+jets normalisation          & 1.1 &    2.1    \\
$t\bar{t}$ modelling            & 0.8 &   1.2  \\
Single-top quark background modelling            & 0.6 &    2.1    \\
$W + \geq1$ light  jets modelling     & 0.3 &  0.4      \\
Diboson normalisation           & 0.1 &    0.3    \\
\midrule
Jet energy resolution      & 4.6 &  7.8 \\
\sxut\ JES correction   & 4.4 &    5.1   \\
Jet energy scale           & 4.0 &  5.3      \\
Flavour tagging            & 2.0 &     1.3  \\
Electron reconstruction    & 1.4 &     0.5   \\
Muon reconstruction        & 1.3 &     0.7  \\
Integrated luminosity                 & 1.3 &  0.4 \\
$E_{\text{T}}^{\text{miss}}$ & 0.6 &    2.4  \\
Jet-vertex tagging                       & 0.07 &    0.05    \\

\midrule
Simulation's statistical uncertainty 	& 2.3 & 6.5 \\
\midrule
Data's statistical uncertainty & 16 & 38  \\
Total systematic uncertainty  & 15 &  18 \\
\midrule
Total uncertainty	      & 21 & 42 \\
\bottomrule
\end{tabular}

\label{tab:SysImpactData}
\end{table}

\begin{figure}[t]
\begin{center}
\subfloat[ ]{
\includegraphics[width=8.2cm,height=8.2cm,keepaspectratio]{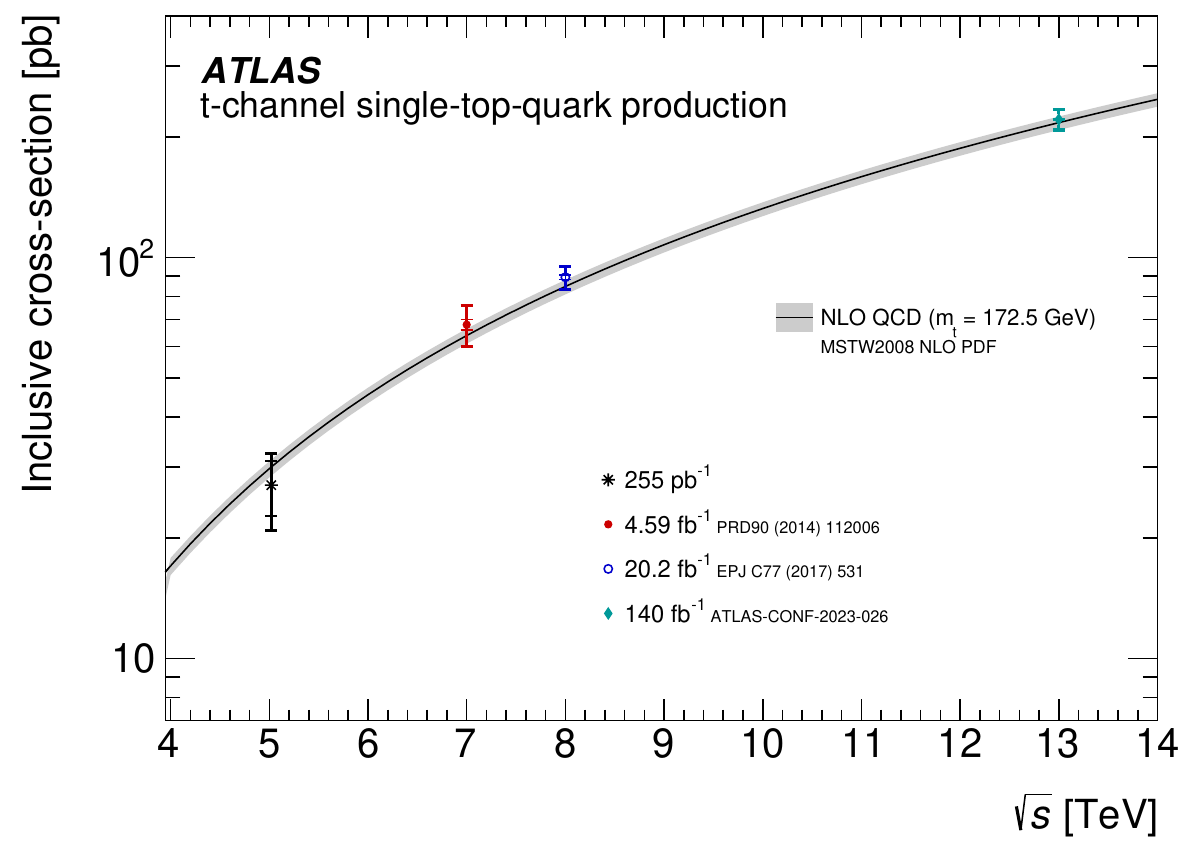}
}
\quad
\subfloat[ ]{
\includegraphics[width=6cm,height=6cm,keepaspectratio]{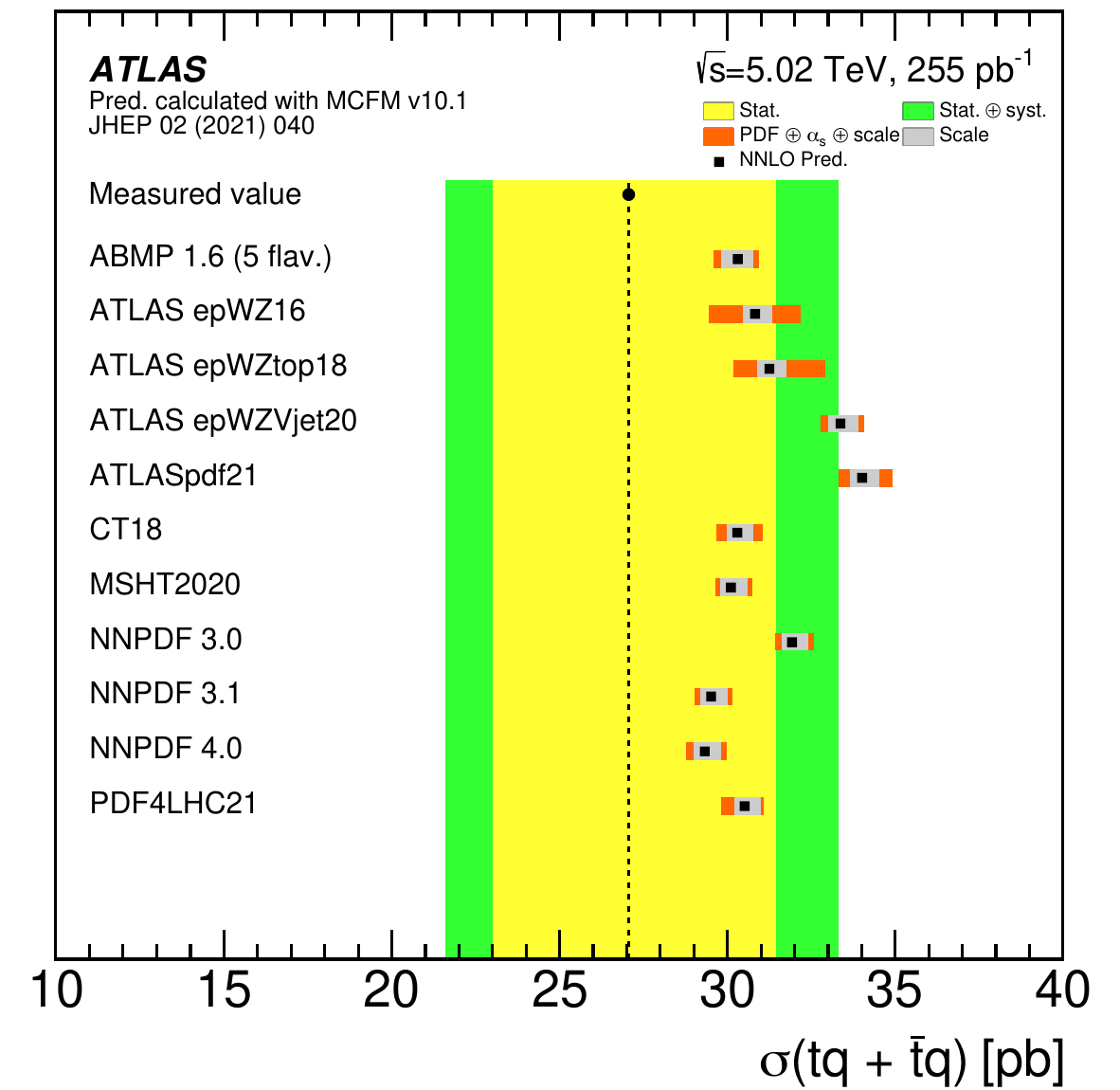}
}

\caption{Summary of (a) ATLAS measurements of the $t$-channel single-top-quark production cross-sections as a function of the centre-of mass energy and (b) the measured \xtchan\ at $\rts = 5.02$ \TeV. In (a), the measurements are compared with theoretical calculations at NLO in QCD~\cite{hathor, hathor15}. In (b), the dashed line and the dot show the measured value, the yellow band displays the statistical uncertainty, and the green band displays the total uncertainty on the measurement. For comparison, the predictions of MCFM based on different PDFs are included. The gray (orange) band represents the uncertainty on the predictions arising from the scale variations (scale, PDF and \alphas\ variations added in quadrature).}
\label{fig:cmeTChanSummary}
\end{center}
\end{figure}

Figure~\ref{fig:cmeTChanSummary} presents a summary of $t$-channel single-top-quark cross-section measurements by the ATLAS Collaboration as a function of the centre-of-mass energy and the measurement at $\rts = 5.02$ \TeV.
The NLO prediction from the \MCFM MC generator is compared with the measured cross-sections
and describes well the evolution of the
single-top-quark production cross-section in the $t$-channel as a function of $\sqrt{s}$.

From the fitted ratio and the total cross-section, the individual single-top-quark and
single-top-antiquark cross-sections are
$\sigma(t q) = \Writetx$~pb and
$\sigma(\bar{t}q) = \Writetbx$~pb.

The $t$-channel single-top-quark production cross-section depends on $\fLV^2 \cdot |V_{tb}|^2$, where \fLV\ is a left-handed form factor that is unity in the SM and $V_{tb}$ is a component of the CKM matrix~\cite{ATLAS:2019hhu}.
By assuming that the CKM matrix elements $|V_{td}|$ and $|V_{ts}|$ are much smaller
than $|V_{tb}|$\ and that the $Wtb$ vertex is left-handed,
the measured cross-section gives $\fLV \cdot |V_{tb}| = \WriteV$. The experimental uncertainties in the measured \xtchan\ and the uncertainties in the predicted \xtchan\ arising from scale variations, choice of PDF, $\alphas$, and $m_t$ dependence are all summed in quadrature.

\FloatBarrier
\section{Conclusions}
The single-top-quark $t$-channel production cross-section is measured at \rtsf\ using
\pp\ data collected with the ATLAS detector corresponding to an integrated luminosity of 255~pb$^{-1}$.
The analysis is performed by selecting semileptonic decays of
the top quark with exactly two jets in the final state, one of which is required to be $b$-tagged.
After performing a profile maximum-likelihood fit to the BDT discriminant distributions in the $\ell^{+}$+jets and $\ell^{-}$+jets channels, the combined single-top-quark and single-top-antiquark production cross-section in the $t$-channel is $\Writetcx$\,pb.

This measurement of single-top-quark production at \rtsf\ is in good agreement with the
SM prediction.
Although its uncertainty is four times larger than measurements at higher centre-of-mass energies,
it provides another independent test of the SM predictions.

The ratio of the single-top-quark and single-top-antiquark production cross-sections is measured to be $R_t = \WriteRt$ and is also in good agreement with the SM\@.
The observed individual single-top-quark and single-top-antiquark production cross-sections are
$\sigma(t q) = \Writetx$~pb and $\sigma(\bar{t}q) = \Writetbx$~pb.
Finally, the product of the left-handed form factor and $V_{tb}$ extracted from the
cross-section measurement at $\rts = 5.02$ \TeV\ is $\WriteV$.


\clearpage
\section*{Acknowledgments}
%

%
%

%
%

We thank CERN for the very successful operation of the LHC and its injectors, as well as the support staff at
CERN and at our institutions worldwide without whom ATLAS could not be operated efficiently.

The crucial computing support from all WLCG partners is acknowledged gratefully, in particular from CERN, the ATLAS Tier-1 facilities at TRIUMF/SFU (Canada), NDGF (Denmark, Norway, Sweden), CC-IN2P3 (France), KIT/GridKA (Germany), INFN-CNAF (Italy), NL-T1 (Netherlands), PIC (Spain), RAL (UK) and BNL (USA), the Tier-2 facilities worldwide and large non-WLCG resource providers. Major contributors of computing resources are listed in Ref.~\cite{ATL-SOFT-PUB-2023-001}.

We gratefully acknowledge the support of ANPCyT, Argentina; YerPhI, Armenia; ARC, Australia; BMWFW and FWF, Austria; ANAS, Azerbaijan; CNPq and FAPESP, Brazil; NSERC, NRC and CFI, Canada; CERN; ANID, Chile; CAS, MOST and NSFC, China; Minciencias, Colombia; MEYS CR, Czech Republic; DNRF and DNSRC, Denmark; IN2P3-CNRS and CEA-DRF/IRFU, France; SRNSFG, Georgia; BMBF, HGF and MPG, Germany; GSRI, Greece; RGC and Hong Kong SAR, China; ISF and Benoziyo Center, Israel; INFN, Italy; MEXT and JSPS, Japan; CNRST, Morocco; NWO, Netherlands; RCN, Norway; MEiN, Poland; FCT, Portugal; MNE/IFA, Romania; MESTD, Serbia; MSSR, Slovakia; ARRS and MIZ\v{S}, Slovenia; DSI/NRF, South Africa; MICINN, Spain; SRC and Wallenberg Foundation, Sweden; SERI, SNSF and Cantons of Bern and Geneva, Switzerland; MOST, Taipei; TENMAK, T\"urkiye; STFC, United Kingdom; DOE and NSF, United States of America.

Individual groups and members have received support from BCKDF, CANARIE, CRC and DRAC, Canada; CERN-CZ, PRIMUS 21/SCI/017 and UNCE SCI/013, Czech Republic; COST, ERC, ERDF, Horizon 2020, ICSC-NextGenerationEU and Marie Sk{\l}odowska-Curie Actions, European Union; Investissements d'Avenir Labex, Investissements d'Avenir Idex and ANR, France; DFG and AvH Foundation, Germany; Herakleitos, Thales and Aristeia programmes co-financed by EU-ESF and the Greek NSRF, Greece; BSF-NSF and MINERVA, Israel; Norwegian Financial Mechanism 2014-2021, Norway; NCN and NAWA, Poland; La Caixa Banking Foundation, CERCA Programme Generalitat de Catalunya and PROMETEO and GenT Programmes Generalitat Valenciana, Spain; G\"{o}ran Gustafssons Stiftelse, Sweden; The Royal Society and Leverhulme Trust, United Kingdom.

In addition, individual members wish to acknowledge support from Chile: Agencia Nacional de Investigaci\'on y Desarrollo (FONDECYT 1190886, FONDECYT 1210400, FONDECYT 1230812, FONDECYT 1230987); China: National Natural Science Foundation of China (NSFC - 12175119, NSFC 12275265, NSFC-12075060); Czech Republic: PRIMUS Research Programme (PRIMUS/21/SCI/017); European Union: European Research Council (ERC - 948254), Horizon 2020 Framework Programme (MUCCA - CHIST-ERA-19-XAI-00), European Union, Future Artificial Intelligence Research (FAIR-NextGenerationEU PE00000013), Italian Center for High Performance Computing, Big Data and Quantum Computing (ICSC, NextGenerationEU), Marie Sklodowska-Curie Actions (EU H2020 MSC IF GRANT NO 101033496); France: Agence Nationale de la Recherche (ANR-20-CE31-0013, ANR-21-CE31-0013, ANR-21-CE31-0022), Investissements d'Avenir Idex (ANR-11-LABX-0012), Investissements d'Avenir Labex (ANR-11-LABX-0012); Germany: Baden-Württemberg Stiftung (BW Stiftung-Postdoc Eliteprogramme), Deutsche Forschungsgemeinschaft (DFG - CR 312/5-1); Italy: Istituto Nazionale di Fisica Nucleare (FELLINI G.A. n. 754496, ICSC, NextGenerationEU); Japan: Japan Society for the Promotion of Science (JSPS KAKENHI JP21H05085, JSPS KAKENHI JP22H01227, JSPS KAKENHI JP22H04944); Netherlands: Netherlands Organisation for Scientific Research (NWO Veni 2020 - VI.Veni.202.179); Norway: Research Council of Norway (RCN-314472); Poland: Polish National Agency for Academic Exchange (PPN/PPO/2020/1/00002/U/00001), Polish National Science Centre (NCN 2021/42/E/ST2/00350, NCN UMO-2019/34/E/ST2/00393, UMO-2020/37/B/ST2/01043, UMO-2021/40/C/ST2/00187); Slovenia: Slovenian Research Agency (ARIS grant J1-3010); Spain: BBVA Foundation (LEO22-1-603), Generalitat Valenciana (Artemisa, FEDER, IDIFEDER/2018/048), La Caixa Banking Foundation (LCF/BQ/PI20/11760025), Ministry of Science and Innovation (MCIN \& NextGenEU PCI2022-135018-2, MICIN \& FEDER PID2021-125273NB, RYC2019-028510-I, RYC2020-030254-I, RYC2021-031273-I, RYC2022-038164-I), PROMETEO and GenT Programmes Generalitat Valenciana (CIDEGENT/2019/023, CIDEGENT/2019/027); Sweden: Swedish Research Council (VR 2018-00482, VR 2022-03845, VR 2022-04683, VR grant 2021-03651), Knut and Alice Wallenberg Foundation (KAW 2017.0100, KAW 2018.0157, KAW 2018.0458, KAW 2019.0447); Switzerland: Swiss National Science Foundation (SNSF - PCEFP2\_194658); United Kingdom: Leverhulme Trust (Leverhulme Trust RPG-2020-004); United States of America: U.S. Department of Energy (ECA DE-AC02-76SF00515), Neubauer Family Foundation.

%
%


%
%
%
\printbibliography

@article{Campbell_2010,
 	author = "Campbell, John M. and Ellis, R. K.",
    editor = {Bl\"umlein, Johannes and Moch, Sven-Olaf and Riemann, Tord},
    title = "{MCFM for the Tevatron and the LHC}",
    eprint = "1007.3492",
    archivePrefix = "arXiv",
    primaryClass = "hep-ph",
    reportNumber = "FERMILAB-CONF-10-244-T",
    doi = "10.1016/j.nuclphysbps.2010.08.011",
    journal = "Nucl. Phys. B Proc. Suppl.",
    volume = "205-206",
    pages = "10--15",
    year = "2010"
}

@article{hathor,
author = "Aliev, M. and Lacker, H. and Langenfeld, U. and Moch, S. and Uwer, P. and Wiedermann, M.",
    title = "{HATHOR -- HAdronic Top and Heavy quarks crOss section calculatoR}",
    doi = "10.1016/j.cpc.2010.12.040",
    journal = "Comput. Phys. Commun.",
    volume = "182",
    pages = "1034--1046",
    year = "2011",
    eprint = "1007.1327",
    archivePrefix = "arXiv",
    primaryClass = "hep-ph",
}

@article{hathor15,
    author = {Kant, P. and Kind, O. M. and Kintscher, T. and Lohse, T. and Martini, T. and M\"olbitz, S. and Rieck, P. and Uwer, P.},
    title = "{HatHor  for single top-quark production: Updated predictions and uncertainty estimates for single top-quark production in hadronic collisions}",
    doi = "10.1016/j.cpc.2015.02.001",
    journal = "Comput. Phys. Commun.",
    volume = "191",
    pages = "74--89",
    year = "2015",
    eprint = "1406.4403",
    archivePrefix = "arXiv",
    primaryClass = "hep-ph",
}

@Article{toppp,
      author         = "Czakon, Michal and Mitov, Alexander",
      title          = "{Top++: A program for the calculation of the top-pair
                        cross-section at hadron colliders}",
      journal        = "Comput.\ Phys.\ Commun.",
      volume         = "185",
      pages          = "2930",
      doi            = "10.1016/j.cpc.2014.06.021",
      year           = "2014",
    eprint = "1112.5675",
    archivePrefix = "arXiv",
    primaryClass = "hep-ph",
}

@Article{Butterworth,
      author         = "Butterworth, Jon and others",
      title          = "{PDF4LHC recommendations for LHC Run II}",
      journal        = "J. Phys. G",
      volume         = "43",
      year           = "2016",
      pages          = "023001",
      doi            = "10.1088/0954-3899/43/2/023001",
    eprint = "1510.03865",
    archivePrefix = "arXiv",
    primaryClass = "hep-ph",
}

@article{Willenbrock:1986cr,
    author = "Willenbrock, Scott S. D. and Dicus, Duane A.",
    title = "{Production of heavy quarks from W-gluon fusion}",
    reportNumber = "UTTG-02-86",
    doi = "10.1103/PhysRevD.34.155",
    journal = "Phys. Rev. D",
    volume = "34",
    pages = "155",
    year = "1986"
}

@article{CDF,
  title = {Observation of Electroweak Single Top-Quark Production},
  author = "{CDF Collaboration}",
  journal = {Phys. Rev. Lett.},
  volume = {103},
  pages = {092002},
  numpages = {8},
  year = {2009},
  publisher = {American Physical Society},
  doi = {10.1103/PhysRevLett.103.092002},
  eprint = "0903.0885",
  archivePrefix = "arXiv",
  primaryClass = "hep-ex",
%  url = {https://link.aps.org/doi/10.1103/PhysRevLett.103.092002}
}

@article{DO,
  title = {Observation of Single Top-Quark Production},
  author = "{D0 Collaboration}",
  journal = {Phys. Rev. Lett.},
  volume = {103},
  pages = {092001},
  numpages = {7},
  year = {2009},
  publisher = {American Physical Society},
  doi = {10.1103/PhysRevLett.103.092001},
  eprint = "0903.0850",
  archivePrefix = "arXiv",
  primaryClass = "hep-ex",
%  url = {https://link.aps.org/doi/10.1103/PhysRevLett.103.092001}
}

@article{PhysRevLett.10.531,
  title = {Unitary Symmetry and Leptonic Decays},
  author = {Cabibbo, Nicola},
  journal = {Phys. Rev. Lett.},
  volume = {10},
  pages = {531--533},
  numpages = {0},
  year = {1963},
  publisher = {American Physical Society},
  doi = {10.1103/PhysRevLett.10.531},
%  url = {https://link.aps.org/doi/10.1103/PhysRevLett.10.531}
}

@article{PTP.49.652,
    author = {Kobayashi, Makoto and Maskawa, Toshihide},
    title = "{CP-Violation in the Renormalizable Theory of Weak Interaction}",
    journal = {Progress of Theoretical Physics},
    volume = {49},
    number = {2},
    pages = {652-657},
    year = {1973},
    month = {02},
 %   issn = {0033-068X},
    doi = {10.1143/PTP.49.652},
 %   url = {https://doi.org/10.1143/PTP.49.652},
 %   eprint = {https://academic.oup.com/ptp/article-pdf/49/2/652/5257692/49-2-652.pdf},
}

@article{Schwienhorst,
    author = "Giammanco, Andrea and Schwienhorst, Reinhard",
    title = "{Single top-quark production at the Tevatron and the LHC}",
    doi = "10.1103/RevModPhys.90.035001",
    journal = "Rev. Mod. Phys.",
    volume = "90",
    number = "3",
    pages = "035001",
    year = "2018",
    eprint = "1710.10699",
    archivePrefix = "arXiv",
    primaryClass = "hep-ex"
}

@article{cttenpdf,
  title = "{New parton distributions for collider physics}",
  author = {Hung-Liang Lai and others},
  journal        = "Phys. Rev. D",
  volume = {82},
  pages = {074024},
  year = {2010},
  doi = {10.1103/PhysRevD.82.074024},
    eprint = "1007.2241",
    archivePrefix = "arXiv",
    primaryClass = "hep-ph",
}

@article{pythia8,
      author         = {Sj\"ostrand, Torbjorn and Mrenna, Stephen and Skands, Peter},
      title          = "{A brief introduction to PYTHIA 8.1}",
      journal        = "Comput. Phys. Commun.",
      volume         = "178",
      pages          = "852",
      doi            = "10.1016/j.cpc.2008.01.036",
      year           = "2008",
    eprint = "0710.3820",
    archivePrefix = "arXiv",
    primaryClass = "hep-ph",
}

@article{sherpa22,
    author = "Bothmann, Enrico and others",
    collaboration = "Sherpa",
    title = "{Event generation with Sherpa 2.2}",
    doi = "10.21468/SciPostPhys.7.3.034",
    journal = "SciPost Phys.",
    volume = "7",
    number = "3",
    pages = "034",
    year = "2019",
    eprint = "1905.09127",
    archivePrefix = "arXiv",
    primaryClass = "hep-ph",
}

@article{zxsecnnlo,
    author = "Anastasiou, Charalampos and Dixon, Lance and Melnikov, Kirill and Petriello, Frank",
    title = "{High-precision QCD at hadron colliders: Electroweak gauge boson rapidity distributions at next-to-next-to-leading order}",
    doi = "10.1103/PhysRevD.69.094008",
    journal = "Phys. Rev. D",
    volume = "69",
    pages = "094008",
    year = "2004",
    eprint = "hep-ph/0312266",
    archivePrefix = "arXiv"
}

@inproceedings{XGBoost,
 author = {Chen, Tianqi and Guestrin, Carlos},
 title = {{XGBoost}: A Scalable Tree Boosting System},
 booktitle = {Proceedings of the 22nd ACM SIGKDD International Conference on Knowledge Discovery and Data Mining},
 series = {KDD '16},
 year = {2016},
 isbn = {978-1-4503-4232-2},
 location = {San Francisco, California, USA},
 pages = {785--794},
 numpages = {10},
 doi = {10.1145/2939672.2939785},
 eprint = "1603.02754",
 archivePrefix = "arXiv",
 primaryClass = "cs.LG",
}

@article{ATLAS:2019hhu,
    author = "{ATLAS and CMS Collaborations}",
    title = "{Combinations of single-top-quark production cross-section measurements and |f$_{LV}$V$_{tb}$| determinations at $ \sqrt{s} $ = 7 and 8 TeV with the ATLAS and CMS experiments}",
    doi = "10.1007/JHEP05(2019)088",
    journal = "JHEP",
    volume = "05",
    pages = "088",
    year = "2019",
    eprint = "1902.07158",
    archivePrefix = "arXiv",
    primaryClass = "hep-ex",
}

@Article{TOPQ-2011-14,
    author         = "{ATLAS Collaboration}",
    title          = "{Measurement of the \(t\)-channel single top-quark production cross section in \(pp\) collisions at \(\sqrt{s} = 7\,\text{TeV}\) with the ATLAS detector}",
    journal        = "Phys. Lett. B",
    volume         = "717",
    year           = "2012",
    pages          = "330",
    doi            = "10.1016/j.physletb.2012.09.031",
    eprint = "1205.3130",
    archivePrefix = "arXiv",
    primaryClass = "hep-ex",
}

@Article{TOPQ-2012-21,
    author         = "{ATLAS Collaboration}",
    title          = "{Comprehensive measurements of \(t\)-channel single top-quark production cross sections at \(\sqrt{s} = 7\,\text{TeV}\) with the ATLAS detector}",
    journal        = "Phys. Rev. D",
    volume         = "90",
    year           = "2014",
    pages          = "112006",
    doi            = "10.1103/PhysRevD.90.112006",
    eprint = "1406.7844",
    archivePrefix = "arXiv",
    primaryClass = "hep-ex",
}

@Article{CMS-TOP-10-008,
    author         = "{CMS Collaboration}",
    title          = "{Measurement of the \(t\)-Channel Single Top Quark Production Cross Section in \(pp\) Collisions at \(\sqrt{s} = 7\,\text{TeV}\)}",
    journal        = "Phys. Rev. Lett.",
    volume         = "107",
    year           = "2011",
    pages          = "091802",
    doi            = "10.1103/PhysRevLett.107.091802",
    eprint = "1106.3052",
    archivePrefix = "arXiv",
    primaryClass = "hep-ex",
 }

@Article{CMS-TOP-11-021,
    author         = "{CMS Collaboration}",
    title          = "{Measurement of the single-top-quark \(t\)-channel cross section in \(pp\) collisions at \(\sqrt{s} = 7\,\text{TeV}\)}",
    journal        = "JHEP",
    volume         = "12",
    year           = "2012",
    pages          = "035",
    doi            = "10.1007/JHEP12(2012)035",
    eprint = "1209.4533",
    archivePrefix = "arXiv",
    primaryClass = "hep-ex", 
}

@Article{TOPQ-2015-05,
    author         = "{ATLAS Collaboration}",
    title          = "{Fiducial, total and differential cross-section measurements of \(t\)-channel single top-quark production in \(pp\) collisions at \(8\,\text{TeV}\) using data collected by the ATLAS detector}",
    journal        = "Eur. Phys. J. C",
    volume         = "77",
    year           = "2017",
    pages          = "531",
    doi            = "10.1140/epjc/s10052-017-5061-9",
    eprint = "1702.02859",
    archivePrefix = "arXiv",
    primaryClass = "hep-ex", 
 }

@Article{CMS-TOP-12-038,
    author         = "{CMS Collaboration}",
    title          = "{Measurement of the \(t\)-channel single-top-quark production cross section and of the \(|V_{tb}|\) CKM matrix element in \(pp\) collisions at \(\sqrt{s} = 8\,\text{TeV}\)}",
    journal        = "JHEP",
    volume         = "06",
    year           = "2014",
    pages          = "090",
    doi            = "10.1007/JHEP06(2014)090",
   eprint = "1403.7366",
    archivePrefix = "arXiv",
    primaryClass = "hep-ex",
}

@Article{TOPQ-2015-15,
    author         = "{ATLAS Collaboration}",
    title          = "{Measurement of the inclusive cross-sections of single top-quark and top-antiquark \(t\)-channel production in \(pp\) collisions at \(\sqrt{s} = 13\,\text{TeV}\) with the ATLAS detector}",
    journal        = "JHEP",
    volume         = "04",
    year           = "2017",
    pages          = "086",
    doi            = "10.1007/JHEP04(2017)086",
    eprint = "1609.03920",
    archivePrefix = "arXiv",
    primaryClass = "hep-ex",
}

@Article{CMS-TOP-16-003,
    author         = "{CMS Collaboration}",
    title          = "{Cross section measurement of \(t\)-channel single top quark production in \(pp\) collisions at \(\sqrt{s} = 13\,\text{TeV}\)}",
    journal        = "Phys. Lett. B",
    volume         = "772",
    year           = "2017",
    pages          = "752",
    doi            = "10.1016/j.physletb.2017.07.047",
    eprint = "1610.00678",
    archivePrefix = "arXiv",
    primaryClass = "hep-ex",
}

@Article{CMS-TOP-17-011,
    author         = "{CMS Collaboration}",
    title          = "{Measurement of the single top quark and antiquark production cross sections in the \(t\) channel and their ratio in proton--proton collisions at \(\sqrt{s} = 13\,\text{TeV}\)}",
    journal        = "Phys. Lett. B",
    volume         = "800",
    year           = "2020",
    pages          = "135042",
    doi            = "10.1016/j.physletb.2019.135042",
    eprint = "1812.10514",
    archivePrefix = "arXiv",
    primaryClass = "hep-ex",
}

@Article{PERF-2007-01,
    author         = "{ATLAS Collaboration}",
    title          = "{The ATLAS Experiment at the CERN Large Hadron Collider}",
    journal        = "JINST",
    volume         = "3",
    year           = "2008",
    pages          = "S08003",
    doi            = "10.1088/1748-0221/3/08/S08003",
}

@Article{TRIG-2016-01,
    author         = "{ATLAS Collaboration}",
    title          = "{Performance of the ATLAS trigger system in 2015}",
    journal        = "Eur. Phys. J. C",
    volume         = "77",
    year           = "2017",
    pages          = "317",
    doi            = "10.1140/epjc/s10052-017-4852-3",
    eprint = "1611.09661",
    archivePrefix = "arXiv",
    primaryClass = "hep-ex",
}

@Booklet{ATL-SOFT-PUB-2021-001,
    author         = "{ATLAS Collaboration}",
    title          = "{The ATLAS Collaboration Software and Firmware}",
    howpublished   = "{ATL-SOFT-PUB-2021-001}",
    url            = "https://cds.cern.ch/record/2767187",
    year           = "2021",
}

@Article{DAPR-2018-01,
    author         = "{ATLAS Collaboration}",
    title          = "{ATLAS data quality operations and performance for 2015--2018 data-taking}",
    journal        = "JINST",
    volume         = "15",
    year           = "2020",
    pages          = "P04003",
    doi            = "10.1088/1748-0221/15/04/P04003",
    eprint = "1911.04632",
    archivePrefix = "arXiv",
    primaryClass = "physics.ins-det",
}

@Booklet{ATLAS-CONF-2020-023,
    author         = "{ATLAS Collaboration}",
    title          = "{Luminosity determination for low-pileup datasets at \(\sqrt{s} = 5\) and \(13\,\text{TeV}\) using the ATLAS detector at the LHC}",
    howpublished   = "{ATLAS-CONF-2020-023}",
    url            = "https://cds.cern.ch/record/2725195",
    year           = "2020",
}

@Article{EGAM-2018-01,
    author         = "{ATLAS Collaboration}",
    title          = "{Electron and photon performance measurements with the ATLAS detector using the 2015--2017 LHC proton--proton collision data}",
    journal        = "JINST",
    volume         = "14",
    year           = "2019",
    pages          = "P12006",
    doi            = "10.1088/1748-0221/14/12/P12006",
    eprint = "1908.00005",
    archivePrefix = "arXiv",
    primaryClass = "hep-ex",
}

@Article{MUON-2018-03,
    author         = "{ATLAS Collaboration}",
    title          = "{Muon reconstruction and identification efficiency in ATLAS using the full Run~2 \(pp\) collision data set at \(\sqrt{s} = 13\,\text{TeV}\)}",
    journal        = "Eur. Phys. J. C",
    volume         = "81",
    year           = "2021",
    pages          = "578",
    doi            = "10.1140/epjc/s10052-021-09233-2",
    eprint = "2012.00578",
    archivePrefix = "arXiv",
    primaryClass = "hep-ex",
 }

@Article{TOPQ-2015-09,
    author         = "{ATLAS Collaboration}",
    title          = "{Measurement of the \(t\bar{t}\) production cross-section using \(e\mu\) events with \(b\)-tagged jets in \(pp\) collisions at \(\sqrt{s} = 13\,\text{TeV}\) with the ATLAS detector}",
    journal        = "Phys. Lett. B",
    volume         = "761",
    year           = "2016",
    pages          = "136",
    doi            = "10.1016/j.physletb.2016.08.019",
    eprint = "1606.02699",
    archivePrefix = "arXiv",
    primaryClass = "hep-ex",
    related        = "TOPQ-2015-09-err",
    relatedstring  = "Erratum:",
}

@Article{PERF-2015-09,
    author         = "{ATLAS Collaboration}",
    title          = "{Jet reconstruction and performance using particle flow with the ATLAS Detector}",
    journal        = "Eur. Phys. J. C",
    volume         = "77",
    year           = "2017",
    pages          = "466",
    doi            = "10.1140/epjc/s10052-017-5031-2",
    eprint = "1703.10485",
    archivePrefix = "arXiv",
    primaryClass = "hep-ex",
 }

@Article{JETM-2018-05,
    author         = "{ATLAS Collaboration}",
    title          = "{Jet energy scale and resolution measured in proton--proton collisions at \(\sqrt{s} = 13\,\text{TeV}\) with the ATLAS detector}",
    journal        = "Eur. Phys. J. C",
    volume         = "81",
    year           = "2021",
    pages          = "689",
    doi            = "10.1140/epjc/s10052-021-09402-3",
    eprint = "2007.02645",
    archivePrefix = "arXiv",
    primaryClass = "hep-ex",
}

@Article{PERF-2014-03,
    author         = "{ATLAS Collaboration}",
    title          = "{Performance of pile-up mitigation techniques for jets in \(pp\) collisions at \(\sqrt{s} = 8\,\text{TeV}\) using the ATLAS detector}",
    journal        = "Eur. Phys. J. C",
    volume         = "76",
    year           = "2016",
    pages          = "581",
    doi            = "10.1140/epjc/s10052-016-4395-z",
    eprint = "1510.03823",
    archivePrefix = "arXiv",
    primaryClass = "hep-ex",
}

@article{ATLAS:2022qxm,
    author         = "{ATLAS Collaboration}",
    title = "{ATLAS flavour-tagging algorithms for the LHC Run 2 $pp$ collision dataset}",
     journal        = "Eur. Phys. J. C",
    volume         = "83",
    year           = "2023",
    pages          = "681",
    doi            = "10.1140/epjc/s10052-023-11699-1",
    eprint = "2211.16345",
    archivePrefix = "arXiv",
    primaryClass = "hep-ex",
}

@Article{PERF-2016-07,
    author         = "{ATLAS Collaboration}",
    title          = "{Performance of missing transverse momentum reconstruction with the ATLAS detector using proton--proton collisions at \(\sqrt{s} = 13\,\text{TeV}\)}",
    journal        = "Eur. Phys. J. C",
    volume         = "78",
    year           = "2018",
    pages          = "903",
    doi            = "10.1140/epjc/s10052-018-6288-9",
    eprint = "1802.08168",
    archivePrefix = "arXiv",
    primaryClass = "hep-ex",
}

@Booklet{ATL-PHYS-PUB-2014-021,
    author         = "{ATLAS Collaboration}",
    title          = "{ATLAS Pythia~8 tunes to \(7~\text{TeV}\) data}",
    howpublished   = "{ATL-PHYS-PUB-2014-021}",
    url            = "https://cds.cern.ch/record/1966419",
    year           = "2014",
}

@Article{TRIG-2018-01,
    author         = "{ATLAS Collaboration}",
    title          = "{Performance of the ATLAS muon triggers in Run~2}",
    journal        = "JINST",
    volume         = "15",
    year           = "2020",
    pages          = "P09015",
    doi            = "10.1088/1748-0221/15/09/p09015",
    eprint = "2004.13447",
    archivePrefix = "arXiv",
    primaryClass = "physics.ins-det",
}

@Article{TRIG-2018-05,
    author         = "{ATLAS Collaboration}",
    title          = "{Performance of electron and photon triggers in ATLAS during LHC Run~2}",
    journal        = "Eur. Phys. J. C",
    volume         = "80",
    year           = "2020",
    pages          = "47",
    doi            = "10.1140/epjc/s10052-019-7500-2",
    eprint = "1909.00761",
    archivePrefix = "arXiv",
    primaryClass = "hep-ex",
}

@Article{PERF-2017-01,
    author         = "{ATLAS Collaboration}",
    title          = "{Electron reconstruction and identification in the ATLAS experiment using the 2015 and 2016 LHC proton--proton collision data at \(\sqrt{s} = 13\,\text{TeV}\)}",
    journal        = "Eur. Phys. J. C",
    volume         = "79",
    year           = "2019",
    pages          = "639",
    doi            = "10.1140/epjc/s10052-019-7140-6",
    eprint = "1902.04655",
    archivePrefix = "arXiv",
    primaryClass = "physics.ins-det",
}

@Article{PERF-2017-03,
    author         = "{ATLAS Collaboration}",
    title          = "{Electron and photon energy calibration with the ATLAS detector using 2015--2016 LHC proton--proton collision data}",
    journal        = "JINST",
    volume         = "14",
    year           = "2019",
    pages          = "P03017",
    doi            = "10.1088/1748-0221/14/03/P03017",
    eprint = "1812.03848",
    archivePrefix = "arXiv",
    primaryClass = "hep-ex",
}

@article{ATLAS:2022jbj,
    author         = "{ATLAS Collaboration}",
    title = "{Measurement of the $t\bar{t}$ production cross-section in $pp$ collisions at $\sqrt{s}=5.02$ TeV with the ATLAS detector}",
    journal        = "JHEP",
    volume         = "06",
    year = "2023",
    pages = "138",
    doi = "https://doi.org/10.1007/JHEP06(2023)138",
    eprint = "2207.01354",
    archivePrefix = "arXiv",
    primaryClass = "hep-ex",    
}

@Article{CMS-SMP-20-012,
    author         = "{CMS Collaboration}",
    title          = "{Measurements of the Electroweak Diboson Production Cross Sections in Proton--Proton Collisions at \(\sqrt{s} = 5.02\,\text{TeV}\) Using Leptonic Decays}",
    journal        = "Phys. Rev. Lett.",
    volume         = "127",
    year           = "2021",
    pages          = "191801",
    doi            = "10.1103/PhysRevLett.127.191801",
    eprint = "2107.01137",
    archivePrefix = "arXiv",
    primaryClass = "hep-ex",
}

@Article{CMS-TOP-20-004,
    author         = "{CMS Collaboration}",
    title          = "{Measurement of the inclusive \(t\bar{t}\) production cross section in proton--proton collisions at \(\sqrt{s} = 5.02\,\text{TeV}\)}",
    journal        = "JHEP",
    volume         = "04",
    year           = "2022",
    pages          = "144",
    doi            = "10.1007/JHEP04(2022)144",
    eprint = "2112.09114",
    archivePrefix = "arXiv",
    primaryClass = "hep-ex",
}

@article{PhysRevLett.102.182003,
  title = {Next-to-Leading-Order Predictions for $t$-Channel Single-Top Production at Hadron Colliders},
  author = {Campbell, J. M. and Frederix, R. and Maltoni, F. and Tramontano, F.},
  journal = {Phys. Rev. Lett.},
  volume = {102},
  pages = {182003},
  numpages = {4},
  year = {2009},
  publisher = {American Physical Society},
  doi = {10.1103/PhysRevLett.102.182003},
  eprint = "0903.0005",
  archivePrefix = "arXiv",
  primaryClass = "hep-ph",
}

@article{PhysRevD.104.094042,
  title = {Testing parton distribution functions with $t$-channel single-top-quark production},
  author = {Campbell, John and Neumann, Tobias and Sullivan, Zack},
  journal = {Phys. Rev. D},
  volume = {104},
  pages = {094042},
  numpages = {6},
  year = {2021},
  publisher = {American Physical Society},
  doi = {10.1103/PhysRevD.104.094042},
    eprint = "2109.10448",
    archivePrefix = "arXiv",
    primaryClass = "hep-ph",
}

@article{Kidonakis:2021vob,
    author = "Kidonakis, Nikolaos and Yamanaka, Nodoka",
    title = "{Higher-order corrections for $tW$ production at high-energy hadron colliders}",
    eprint = "2102.11300",
    archivePrefix = "arXiv",
    primaryClass = "hep-ph",
    doi = "10.1007/JHEP05(2021)278",
    journal = "JHEP",
    volume = "05",
    pages = "278",
    year = "2021"
}

@article{campbell2015multithreaded,
    author = "Campbell, John M. and Ellis, R. Keith and Giele, Walter T.",
    title = "{A multi-threaded version of MCFM}",
    eprint = "1503.06182",
    archivePrefix = "arXiv",
    primaryClass = "physics.comp-ph",
    doi = "10.1140/epjc/s10052-015-3461-2",
    journal = "Eur. Phys. J. C",
    volume = "75",
    pages = "246",
    year = "2015"
}

@article{Campbell:2019,
    author = "Campbell, John and Neumann, Tobias",
    title = "{Precision phenomenology with MCFM}",
    eprint = "1909.09117",
    archivePrefix = "arXiv",
    primaryClass = "hep-ph",
    reportNumber = "FERMILAB-PUB-19-477-T, IIT-CAPP-19-03",
    doi = "10.1007/JHEP12(2019)034",
    journal = "JHEP",
    volume = "12",
    pages = "034",
    year = "2019"
}

@article{ATLAS:2022swp,
    author         = "{ATLAS Collaboration}",
    title = "{Tools for estimating fake/non-prompt lepton backgrounds with the ATLAS detector at the LHC}",
    eprint = "2211.16178",
    archivePrefix = "arXiv",
    primaryClass = "hep-ex",
    reportNumber = "CERN-EP-2022-214",
    month = "11",
    year = "2022"
}

@article{Berends:1990ax,
    author = "Berends, Frits A. and Kuijf, H. and Tausk, B. and Giele, W. T.",
    title = "{On the production of a W and jets at hadron colliders}",
    reportNumber = "FERMILAB-PUB-90-213-T",
    doi = "10.1016/0550-3213(91)90458-A",
    journal = "Nucl. Phys. B",
    volume = "357",
    pages = "32--64",
    year = "1991"
}

@Article{Cacciari:2008gp,
     author    = "Cacciari, Matteo and Salam, Gavin P. and Soyez, Gregory",
     title     = "{The anti-\(k_{t}\) jet clustering algorithm}",
     journal   = "JHEP",
     volume    = "04",
     year      = "2008",
     pages     = "063",
     eprint    = "0802.1189",
     archivePrefix = "arXiv",
     primaryClass  =  "hep-ph",
     doi       = "10.1088/1126-6708/2008/04/063",
     SLACcitation  = "%%CITATION = 0802.1189;%%"
}

@Article{Fastjet,
      author         = "Cacciari, Matteo and Salam, Gavin P. and Soyez, Gregory",
      title          = "{FastJet user manual}",
      journal        = "Eur. Phys. J. C",
      volume         = "72",
      year           = "2012",
      pages          = "1896",
      doi            = "10.1140/epjc/s10052-012-1896-2",
      eprint         = "1111.6097",
      archivePrefix  = "arXiv",
      primaryClass   = "hep-ph",
      reportNumber   = "CERN-PH-TH-2011-297",
      SLACcitation   = "%%CITATION = ARXIV:1111.6097;%%"
}

@article{Ball:2012cx,
	author= {{NNPDF Collaboration} and {R. D. Ball} and {others}},
      title          = "{Parton distributions with LHC data}",
      journal        = "Nucl. Phys. B",
      volume         = "867",
      year           = "2013",
      pages          = "244",
      doi            = "10.1016/j.nuclphysb.2012.10.003",
      eprint         = "1207.1303",
      archivePrefix  = "arXiv",
      primaryClass   = "hep-ph",
      reportNumber   = "EDINBURGH-2012-08, IFUM-FT-997, FR-PHENO-2012-014,
                        RWTH-TTK-12-25, CERN-PH-TH-2012-037, SFB-CPP-12-47\,
                        --CERN-PH-TH-2012-037",
      SLACcitation   = "%%CITATION = ARXIV:1207.1303;%%"
}

@article{Frederix:2012dh,
      author         = "Frederix, Rikkert and Re, Emanuele and Torrielli, Paolo",
      title          = "{Single-top \(t\)-channel hadroproduction in the four-flavour
                        scheme with POWHEG and aMC@NLO}",
      journal        = "JHEP",
      volume         = "09",
      year           = "2012",
      pages          = "130",
      doi            = "10.1007/JHEP09(2012)130",
      eprint         = "1207.5391",
      archivePrefix  = "arXiv",
      primaryClass   = "hep-ph",
      reportNumber   = "ZU-TH-14-12, IPPP-12-54, DCPT-12-108,
                        CERN-PH-TH-2012-206, MCNET-12-10, LPN12-084",
      SLACcitation   = "%%CITATION = ARXIV:1207.5391;%%"
}

@article{Ball:2014uwa,
      author= {{NNPDF Collaboration} and {R. D. Ball} and {others}},
      title          = "{Parton distributions for the LHC run II}",
      collaboration  = "NNPDF",
      journal        = "JHEP",
      volume         = "04",
      year           = "2015",
      pages          = "040",
      doi            = "10.1007/JHEP04(2015)040",
      eprint         = "1410.8849",
      archivePrefix  = "arXiv",
      primaryClass   = "hep-ph",
      reportNumber   = "EDINBURGH-2014-15, IFUM-1034-FT, CERN-PH-TH-2013-253,
                        OUTP-14-11P, CAVENDISH-HEP-14-11",
      SLACcitation   = "%%CITATION = ARXIV:1410.8849;%%"
}

@article{Agostinelli:2002hh,
      author         = "{GEANT4 Collaboration} and Agostinelli, S. and others",
      title          = "{\textsc{Geant4} -- a simulation toolkit}",
      journal        = "Nucl. Instrum. Meth. A",
      volume         = "506",
      year           = "2003",
      pages          = "250",
      doi            = "10.1016/S0168-9002(03)01368-8",
      reportNumber   = "SLAC-PUB-9350, FERMILAB-PUB-03-339",
      SLACcitation   = "%%CITATION = NUIMA,A506,250;%%"
}

@Article{Cowan:2010js,
   Author = {Cowan, Glen and Cranmer, Kyle and Gross, Eilam and Vitells, Ofer},
   Title = {{Asymptotic formulae for likelihood-based tests of new physics}},
   Journal = {Eur. Phys. J. C},
   Volume = {71},
   Year = {2011},
   Pages = {1554},
   doi = "10.1140/epjc/s10052-011-1554-0",
   Eprint = "1007.1727",
   Archiveprefix = {arXiv},
   Primaryclass = {physics.data-an},
   related = "Cowan:2010js-err",
   relatedstring  = "Erratum:",
}

@article{CDF:2015gsg,
    author = "Aaltonen, Timo Antero and others",
    collaboration = "CDF, D0",
    title = "{Tevatron Combination of Single-Top-Quark Cross Sections and Determination of the Magnitude of the Cabibbo-Kobayashi-Maskawa Matrix Element $\bf V_{tb}$}",
    eprint = "1503.05027",
    archivePrefix = "arXiv",
    primaryClass = "hep-ex",
    reportNumber = "FERMILAB-PUB-15-088-E",
    doi = "10.1103/PhysRevLett.115.152003",
    journal = "Phys. Rev. Lett.",
    volume = "115",
    number = "15",
    pages = "152003",
    year = "2015"
}

@Booklet{ATL-SOFT-PUB-2023-001,
    author         = "{ATLAS Collaboration}",
    title          = "{ATLAS Computing Acknowledgements}",
    howpublished   = "{ATL-SOFT-PUB-2023-001}",
    url            = "https://cds.cern.ch/record/2869272",
    year           = 2023
}
\clearpage

 
\begin{flushleft}
\hypersetup{urlcolor=black}
{\Large The ATLAS Collaboration}

\bigskip

\AtlasOrcid[0000-0002-6665-4934]{G.~Aad}$^\textrm{\scriptsize 102}$,
\AtlasOrcid[0000-0002-5888-2734]{B.~Abbott}$^\textrm{\scriptsize 120}$,
\AtlasOrcid[0000-0002-1002-1652]{K.~Abeling}$^\textrm{\scriptsize 55}$,
\AtlasOrcid[0000-0001-5763-2760]{N.J.~Abicht}$^\textrm{\scriptsize 49}$,
\AtlasOrcid[0000-0002-8496-9294]{S.H.~Abidi}$^\textrm{\scriptsize 29}$,
\AtlasOrcid[0000-0002-9987-2292]{A.~Aboulhorma}$^\textrm{\scriptsize 35e}$,
\AtlasOrcid[0000-0001-5329-6640]{H.~Abramowicz}$^\textrm{\scriptsize 151}$,
\AtlasOrcid[0000-0002-1599-2896]{H.~Abreu}$^\textrm{\scriptsize 150}$,
\AtlasOrcid[0000-0003-0403-3697]{Y.~Abulaiti}$^\textrm{\scriptsize 117}$,
\AtlasOrcid[0000-0002-8588-9157]{B.S.~Acharya}$^\textrm{\scriptsize 69a,69b,m}$
% ...
\bigskip
\\

% ...
$^{29}$Physics Department, Brookhaven National Laboratory, Upton NY; United States of America.\\
% ...
$^{35}$$^{(a)}$Facult\'e des Sciences Ain Chock, R\'eseau Universitaire de Physique des Hautes Energies - Universit\'e Hassan II, Casablanca;$^{(b)}$Facult\'{e} des Sciences, Universit\'{e} Ibn-Tofail, K\'{e}nitra;$^{(c)}$Facult\'e des Sciences Semlalia, Universit\'e Cadi Ayyad, LPHEA-Marrakech;$^{(d)}$LPMR, Facult\'e des Sciences, Universit\'e Mohamed Premier, Oujda;$^{(e)}$Facult\'e des sciences, Universit\'e Mohammed V, Rabat;$^{(f)}$Institute of Applied Physics, Mohammed VI Polytechnic University, Ben Guerir; Morocco.\\
% ...
$^{49}$Fakult\"{a}t Physik , Technische Universit{\"a}t Dortmund, Dortmund; Germany.\\
% ...
$^{55}$II. Physikalisches Institut, Georg-August-Universit\"{a}t G\"ottingen, G\"ottingen; Germany.\\
% ...
$^{69}$$^{(a)}$INFN Gruppo Collegato di Udine, Sezione di Trieste, Udine;$^{(b)}$ICTP, Trieste;$^{(c)}$Dipartimento Politecnico di Ingegneria e Architettura, Universit\`a di Udine, Udine; Italy.\\
% ...
$^{102}$CPPM, Aix-Marseille Universit\'e, CNRS/IN2P3, Marseille; France.\\
% ...
$^{117}$Department of Physics, New York University, New York NY; United States of America.\\
% ...
$^{120}$Homer L. Dodge Department of Physics and Astronomy, University of Oklahoma, Norman OK; United States of America.\\
% ...
$^{150}$Department of Physics, Technion, Israel Institute of Technology, Haifa; Israel.\\
$^{151}$Raymond and Beverly Sackler School of Physics and Astronomy, Tel Aviv University, Tel Aviv; Israel.\\
% ...
$^{a}$ Also Affiliated with an institute covered by a cooperation agreement with CERN.\\
$^{b}$ Also at An-Najah National University, Nablus; Palestine.\\
$^{c}$ Also at Borough of Manhattan Community College, City University of New York, New York NY; United States of America.\\
$^{d}$ Also at Center for High Energy Physics, Peking University; China.\\
$^{e}$ Also at Center for Interdisciplinary Research and Innovation (CIRI-AUTH), Thessaloniki; Greece.\\
$^{f}$ Also at Centro Studi e Ricerche Enrico Fermi; Italy.\\
% ...
$^{m}$ Also at Department of Physics, King's College London, London; United Kingdom.\\
% ...

\end{flushleft}


%

%
%

%

%
%
%
%

\end{document}